\documentclass[aps,twocolumn,pra,tightenlines,showpacs]{revtex4}
\usepackage{amsmath,amsthm,amsfonts,amssymb,amsbsy,mathrsfs}
\usepackage{units,times,float,graphicx,bm,comment}
\usepackage[colorlinks,citecolor=blue,linkcolor=blue,hyperindex,dvipdfm]{hyperref}
\begin{document}
\title{Quantum speed limit of a single atom in a squeezed optical cavity mode}
\author{Ya-Jie Ma$^{1}$, Xue-Chen Gao$^{1}$, Shao-Xiong Wu$^{1}$\footnote{sxwu@nuc.edu.cn}, and Chang-shui Yu$^{2}$\footnote{ycs@dlut.edu.cn}}
\affiliation{$^{1}$ School of Semiconductor and Physics, North University of China, Taiyuan 030051, China\\
 $^{2}$ School of Physics, Dalian University of Technology, Dalian 116024, China}
\date{\today }
\begin{abstract}
We theoretically study the quantum speed limit of a single atom trapped in a Fabry-Perot microresonator. The cavity mode will be squeezed when a driving laser is applied to the second-order nonlinear medium, and the effective Hamiltonian can be obtained under the Bogoliubov squeezing transformation. The analytical expression of evolved atom state can be obtained by using the non-Hermitian Schr\"{o}dinger equation for the initial excited state, and the quantum speed limit time coincides very well for both the analytical expression and the master equation method. From the perspective of quantum speed limit, it is more conducive to accelerate the evolution of the quantum state for the large detuning, strong driving and coupling strength. For the initial superposition state case, the form of initial state has more influence on the evolution speed. The quantum speed limit time is not only dependent on the system parameters but also determined by the initial state.

\vspace{12pt}\noindent{\textbf{Keywords:} Quantum speed limit; Squeezing mode; Non-Hermitian Schr\"{o}dinger equation; Master equation}
\end{abstract}
\maketitle

\section{Introduction}
What is the shortest time for a quantum state to evolve to an orthogonal state under unitary evolution? Mandelstam and Tamm gave a satisfactory explanation \cite{Mandelstam1945}. The shortest time $\Delta t$ was interpreted as the intrinsic time scale between two pure states during unitary dynamical evolution, which is called the MT quantum speed limit time and originated from the Heisenberg energy-time uncertainty relation $\Delta t\geq\hbar/(2\Delta E)$. According to the transition probability between two pure states, Margolus and Levitin \cite{Margolus1998} reinvestigated the quantum state evolution and found that the minimum evolution time is related to the mean value of energy $\langle E\rangle$, which is called the ML quantum speed limit time. In order to character the dynamics of the quantum state, the unified quantum speed limit time including MT-type and ML-type is expressed as $t_{\text{qsl}}=\text{max}\{ \pi\hbar/(2\Delta E),\pi\hbar/(2\langle E\rangle)\}$. The quantum speed limit time for closed systems had also been investigated by Anandan and Aharonov \cite{Anandan1990PRL}, Fleming \cite{Fleming1973}, Bhattacharyya \cite{Bhattacharyya1983}, Vaidman \cite{Vaidman1992}, and so on.

The evolution of quantum systems is of great significance in quantum optimal control \cite{Caneva2009PRL,Hegerfeldt2013PRL,Campbell2017PRL,Frank2016}, quantum metrology \cite{Giovannetti2011,Chin2012PRL}, quantum thermodynamics \cite{Deffner2010PRL}, quantum phase transition \cite{Greiner2002Nature,Sachdev book}, and so on (See the comprehensive review \cite{Deffner2017JPA} and references therein). However, the quantum system will inevitably interact with the environment in real quantum information processing, so we have to use the open system approach to deal with decay and decoherence during evolution. Due to the theoretical significance, the concept of quantum speed limit time have been extended to the open quantum system by using the relative purity \cite{del Campo2013PRL} and quantum Fisher information \cite{Taddei2013PRL}, respectively. Employing the Bures angle as the ``distance" measure between two states in Ref. \cite{Deffner2013PRL}, Deffner and Lutz extended and obtained a unified expression of MT-type and ML-type quantum speed limit time from closed systems to open quantum systems.  Inspired by these seminal works, the quantum speed limit for the open quantum system was further investigated widely in variety of methods \cite{Zhang2014,Xu2014,wusx2018,wusx2020,Zhang2018,sun2015,Liu2016,song2016,wei2016,cai2017,huang2022cpb,
lin2022cpb,zhang2015,xu2019,Yu2018,O'Connor2021,Mondal2016,Jones2010,Uzdin,Teittinen2019,Ektesabi2017,
Deffner2017,Diaz2020,Burgarth2022,cheng2022,Mohan2022,Ness2022PRL,Liu2017,Funo2019,Hu2020,Vu2021PRL,
liu2015,Levitin2009PRL,Campaioli2018PRL,Wusx2015,Mirkin2016,Marvian2016,Poggi2019,Du2021,Tian2019}, the speed limit in classical systems \cite{Shiraishi2018PRL,Okuyama2018PRL,Shanahan2018PRL,wucpb2020,Garcia-Pintos2022}, operational definition  of quantum speed limit \cite{Liu2021,Shao2020} and the relation with the gauge-invariant distance \cite{Sun2019PRL,Sun2021PRL} were also be investigated.

The cavity quantum electrodynamics (QED) system \cite{scully1997} is a powerful tool to investigate the interaction between light and matters, such as the Purcell effect \cite{purcell1946,Bloembergen1954}, the photon blockade \cite{Birnbaum2005,Hamsen2017}, the optical non-reciprocity \cite{Yang2019PRL,Yang2019}, and so on. It requires a high-quality resonance factor and a small mode volume to realize the strong coupling strength between the trapped atom and the cavity mode. For the relatively weak coupling cavity, the coupling strength can be amplified by the squeezed field, and the purpose of strong coupling can be achieved \cite{Lv2015PRL, Qin2018PRL}. This mechanism was also applied to enhance the dipole interaction between two atoms in a cavity mode \cite{Wang2019}. This paper will consider the quantum speed limit of a two-level atom trapped in a high-fineness Fabry-Perot microresonator with a second-order nonlinear medium. A classical coherent laser is applied to the nonlinear medium. Utilizing the Bogoliubov squeezing transformation, the effective Hamiltonian is obtained. Meanwhile, the analytical evolved state is acquired by solving the non-Hermitian Schr\"{o}dinger equation for the initial excited state. The quantum speed limit time based on the analytical solution and the master equation coincides very well. The quantum speed limit time is studied based on the master equation for the initial arbitrary superposition state. One can find that the acceleration of the quantum speed limit is not only determined by the system parameters but also related to the initial state.

The structure of this paper is as follows. In Sec. \ref{sec2}, we give the model and the concept of the quantum speed limit. In Sec. \ref{sec3}, for the initial excited state, the analytical solution of the evolved atom state using the non-Hermitian Schr\"{o}dinger equation and the master equation in the squeezed picture is obtained, and the quantum speed limit time is studied. In Sec. \ref{sec4}, the quantum speed limit time for initial superposition states are investigated. The conclusion and discussion are given in Sec. \ref{sec5}.

\section{The Model and the concept of quantum speed limit}\label{sec2}
\begin{figure}
\centering
\includegraphics[width=0.8\columnwidth]{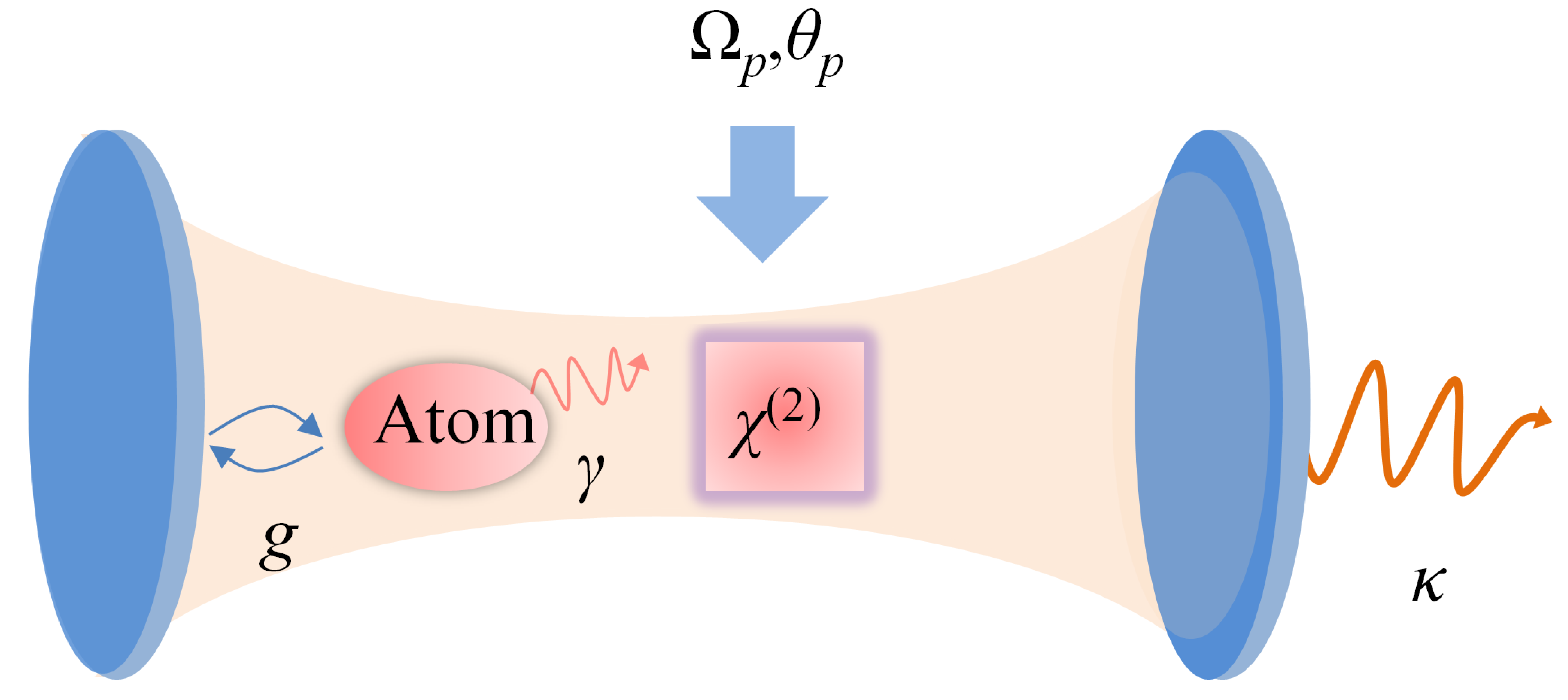}
\caption{The schematic of the model. A two-level atom and a second-order nonlinear medium $\chi^{(2)}$ are trapped in a single-mode optical cavity. A classical coherent driving laser pumps the medium at frequency $\omega_{p}$, amplitude $\Omega_{p}$, and relative phase $\theta_{p}$. $g$ is the coupling strength between the atom and cavity mode, $\gamma$ is the spontaneous emission rate of the atom, and $\kappa$ is the decay rate of the cavity.}\label{fig:1}
\end{figure}

Our model is depicted in Fig. \ref{fig:1}, which consists of a two-level atom and a second-order nonlinear medium trapped in a single-mode high-fineness Fabry-Perot microresonator. When a classical coherent laser drives the nonlinear medium with frequency $\omega_{p}$, amplitude $\Omega_{p}$ and relative phase $\theta_{p}$, the total Hamiltonian of the quantum system and environment is
\begin{align}
H_0= &\omega_a\sigma_{+}\sigma_{-}+\omega_{c}a^{\dagger}a+g(\sigma_{+}a+a^{\dagger}\sigma_{-})\notag\\
&+\frac{\Omega_p}{2}[a^{\dagger2}e^{-i(\theta_p+\omega_pt)}+a^{2}e^{i(\theta_p+\omega_pt)}],\label{H1}
\end{align}
where $\sigma_{+}=|e\rangle\langle g|$ is the raising operator between the ground state $|g\rangle$ and the excited state $|e\rangle$ with transition frequency $\omega_a$, $\omega_c$ is the frequency of bare cavity mode, $a$ is the annihilation operator of the cavity mode, and $g$ is the coupling strength between the single atom and cavity. Under the rotating frame $H_R=U^{\dagger}H_0U-i\hbar U^{\dagger}\dot{U}$, the Hamiltonian (\ref{H1}) is simplified as
\begin{align}
H_R=&\Delta_a\sigma_{+}\sigma_{-}+\Delta_{c}a^{\dagger}a+g(\sigma_{+}a+a^{\dagger}\sigma_{-})\notag\\
 & +\frac{\Omega_p}{2}[e^{i\theta_{p}}a^{2}+e^{-i\theta_{p}}a^{\dagger2}],\label{eq:2}
\end{align}
where the transformation is $U=\exp[-i\omega_l(a^{\dagger}a+\sigma_+\sigma_-)t]$ with the frequency $\omega_{l}=\omega_{p}/2$, $\Delta_{a}=\omega_{a}-\omega_{p}/2$ is the detuning between the driving laser frequency and atom transition frequency, and $\Delta_{c}=\omega_{c}-\omega_{p}/2$ is the detuning between the driving laser frequency and the bare cavity mode frequency. Similar to Refs. \cite{Lv2015PRL,Qin2018PRL,Wang2019}, utilizing the Bogoliubov squeezing transformation $a=a_{s}\cosh(r_{p})-e^{i\theta_{p}}\sinh(r_{p})a_{s}^{\dagger}$ \cite{scully1997}, the cavity mode can be squeezed, and the Hamiltonian (\ref{eq:2}) can be diagonalized in the squeezed picture, where the controllable squeezing parameter $r_{p}$ is  defined by $r_{p}=(1/2)\text{arctanh}\beta$ with $\beta=\Omega_{p}/\Delta_{c}$. By adjusting the relative phase $\theta_{p}$ of the driving laser to be zero and using the rotating-wave approximation, the Hamiltonian (\ref{eq:2}) can be rewritten as
\begin{equation}
H=\Delta_a\sigma_{+}\sigma_{-}+\Delta_{s}a_{s}^{\dagger}a_s
+g_s(\sigma_{+}a_{s}+a_{s}^{\dagger}\sigma_{-}),\label{eq:3}
\end{equation}
where $\Delta_{s}=\Delta_{c}\sqrt{1-\beta^{2}}$ is the cavity detuning in the squeezed picture, $a_s$ denotes the squeezed cavity mode, and $g_s=g\cosh(r_p)$ means the coupling strength between the atom and the squeezed mode $a_s$. When the driving laser is applied, the coupling strength $g_s$ between the atom and the squeezed mode will be exponentially enhanced and transformed into a relatively strong coupling regime. This paper will investigate the quantum speed limit time of the atom's evolution based on this model.

In the seminal work about the quantum speed limit of open quantum system \cite{Deffner2013PRL}, the ML-type quantum speed limit time is given through the von Neumann trace inequality by using the operator norm and trace norm, and the MT-type quantum speed limit time can be obtained by using the Cauchy-Schwarz inequality and Hilbert-Schmidt norm. The unified quantum speed limit time is expressed as
\begin{align}
t_{\text{qsl}}=\text{max}\left\{ \frac{1}{\varLambda_{t}^{\text{op}}},\frac{1}{\varLambda_{t}^{\text{tr}}},
\frac{1}{\varLambda_{t}^{\text{hs}}}\right\} \sin^{2}\{\mathcal{B}[\rho(0),\rho(t)]\},\label{qsl}
\end{align}
where $\varLambda_{t}^{\text{op},\text{tr},\text{hs}}=(1/t)\int_{0}^{t}\text{d}\tau$$\left\Vert \dot{\rho}(t)\right\Vert _{\text{op},\text{tr},\text{hs}}$ with $||\cdot|| _{\text{op},\text{tr},\text{hs}}$ being the operator norm, trace norm and Hilbert-Schmidt norm of the matrix, respectively. The time-dependent non-unitary dynamical operator $\dot{\rho}(t)=$$L_{t}[\rho(t)]$ can be given through the master equation. For the pure initial state $\rho(0)=|\psi(0)\rangle\langle\psi(0)|$ and the final state $\rho(t)$, the Bures angle is defined as $\mathcal{B}[\rho(0),\rho(t)]=\arccos(\sqrt{\langle \psi(0)|\rho(t)|\psi(0)\rangle})$.

\section{The analytical solution of quantum speed limit time}\label{sec3}
\begin{figure}
\centering
\includegraphics[width=1\columnwidth]{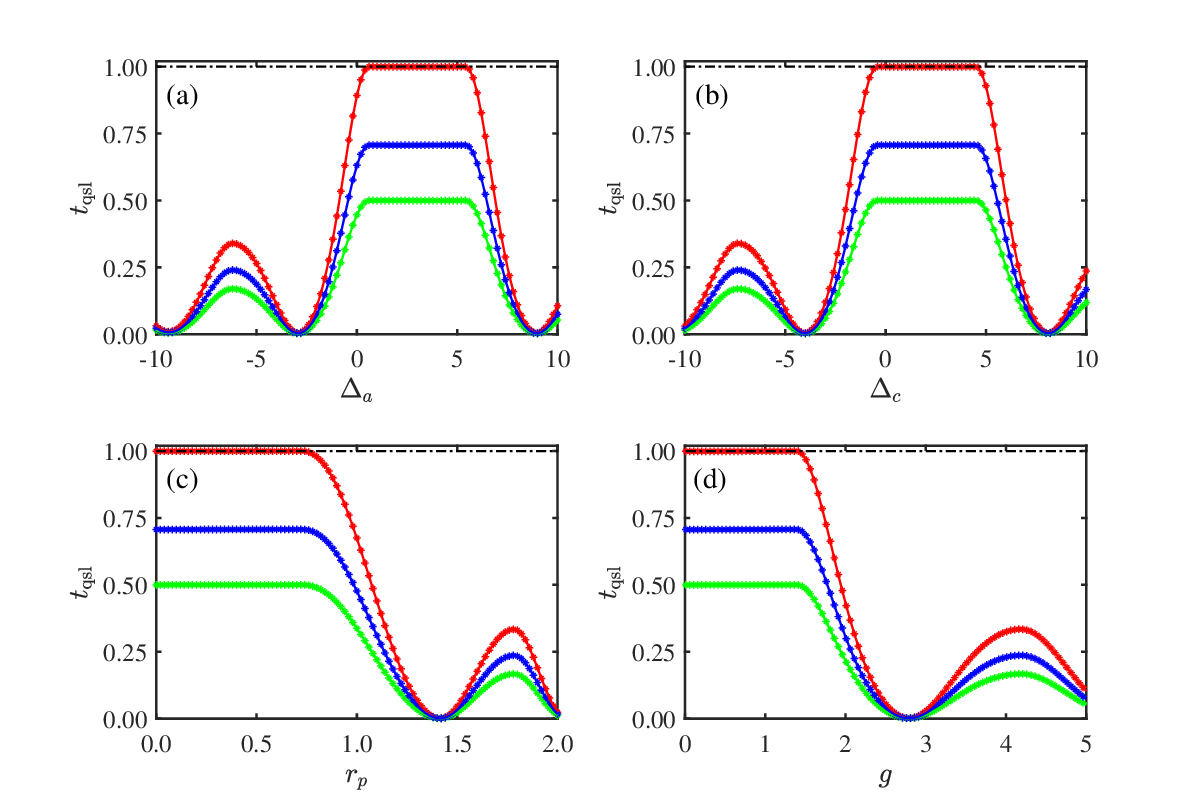}
\caption{The quantum speed limit time $t_{\text{qsl}}$ as functions of the atom detuning $\Delta_{a}$ (a), the cavity detuning $\Delta_{c}$ (b), the squeezing parameter $r_{p}$ (c), the coupling strength $g$ (d). The parameters are $g=1$, $r_{p}=0.1$, and $\Delta_{c}=3g_s/\sqrt{1-\beta^{2}}$ in Panel (a), and $\Delta_{a}=2g$ in Panel (b); The parameters are $\Delta_{c}=3g_s/\sqrt{1-\beta^{2}}$, $\Delta_{a}=2g$, and  $g=1$ in Panel (c), and $r_{p}=0.1$ in Panel (d). The system's initial state is assumed as $\left|e,0\right\rangle $. The solid line is the analytic expression (\ref{qsljiexi}) based on the operator norm (red curve), the Hilbert-Schmidt norm (blue curve), and the trace norm (green curve), respectively. The corresponding asterisk is obtained from the master equation (\ref{qslshuzhi}). In all Panels, the decay rates are $\kappa=10^{-3}g$, $\gamma=10^{-3}g$, and the actual evolution time $\tau=1$ (in units of $g$, black dot dotted line). }\label{fig:2}
\end{figure}
In this section, we consider that the trapped atom is prepared in the excited state and investigate its dynamical evolution. The cavity is assumed as vacuum, $\emph{i.e.}$, the initial state is $|\psi(0)\rangle =|e,0\rangle$, the evolved state at time $t$ has the following form:
\begin{align}
|\psi(t)\rangle =A(t)|e,0\rangle+B(t)|g,1\rangle+C(t)|g,0\rangle .\label{eq:5}
\end{align}
The final target state can be proved according to the reachable state set and Bures angle proposed in Refs. \cite{Liu2021,Shao2020}. Taking into account the spontaneous emission of the atom and the decay of the cavity, the evolved state can be solved by the non-Hermitian Schr\"{o}dinger equation
\begin{equation}
i\frac{\partial|\psi(t)\rangle }{\partial t}=\left(H-\frac{i\gamma}{2}a_{s}^{\dagger}a_{s}-\frac{i\kappa}{2}\sigma_{+}\sigma_{-}\right)
|\psi(t)\rangle,\label{eq:6}
\end{equation}
where $\gamma$ is the atomic spontaneous emission rate, and $\kappa$ describes the bare cavity decay rate. Taking the state (\ref{eq:5}) into the non-Hermitian Schr\"{o}dinger equation (\ref{eq:6}), the coefficients $A(t)$, $B(t)$, and $C(t)$ are determined by the differential equations
\begin{align}
i\dot{A}(t)= & \left(\Delta_{a}-\frac{i\gamma}{2}\right)A(t)+g_sB(t),\notag \\
i\dot{B}(t)= & \left(\Delta_{s}-\frac{i\kappa}{2}\right)B(t)+g_sA(t),\notag\\
i\dot{C}(t)= & 0.
\end{align}
By tedious calculations, the coefficients $A(t)$ and $B(t)$ can be solved analytically as follows:
\begin{align}
A(t)= & \exp\left(-\frac{1}{4}\nu t\right)\left[\cosh\left(\frac{1}{4}\xi t\right)-\frac{\mu}{\xi}\sinh\left(\frac{1}{4}\xi t\right)\right],\notag\\
B(t)= & \frac{4g_s}{i\xi}\exp\left(-\frac{1}{4}\nu t\right)\sinh\left(\frac{1}{4}\xi t\right),
\end{align}
where $\mu=\gamma-\kappa+2i(\Delta_{a}-\Delta_{s})$, $\nu=\gamma+\kappa+2i(\Delta_{a}+\Delta_{s})$, and $\xi=\sqrt{\mu^{2}-16g_s^{2}}$. The coefficient $C(t)$ is a tiny constant quantity related to the decay rates $\gamma,~\kappa$, and the evolution time $t$. Under the specific system parameters, the probability $|C(t)|^2$ is tiny and ignored, and the solution of the state (\ref{eq:5}) is not affected. When we only consider the dynamics of the atom and trace out the influence of the cavity mode, the reduced density matrix operator reads
\begin{equation}
\rho(t)=\left(\begin{array}{cc}
\left|A(t)\right|^{2} & A(t)C^{*}(t)\\
A^{*}(t)C(t) & \left|B(t)\right|^{2}+\left|C(t)\right|^{2}
\end{array}\right).\label{qsljiexi}
\end{equation}
To verify the validity of the analytical calculation, we also calculate the reduced density matrix numerically using the Lindblad master equation in the squeezed picture as a contrast, which is given by
\begin{equation}
\dot{\rho}(t)=i[\rho(t),H]
-\frac{1}{2}\mathcal{L}(L_{\sigma_-})\rho(t)-\frac{1}{2}\mathcal{L}(L_{as})\rho(t).\label{qslshuzhi}
\end{equation}
The detailed derivation of Eq. (\ref{qslshuzhi}) is given in the Appendix. The cavity has been considered coupled to a broadband squeezed vacuum field, so the noise generated by driving laser $\Omega_p$ will be offset by appropriate parameter selection. Both the analytical and numerical methods will be employed to research the quantum speed limit time of the atom in the remainder of this section.

The variation of the quantum speed limit time along with the system parameters $\Delta_a$, $\Delta_c$, $r_p$, and $g$ are described in Fig. \ref{fig:2}, where the solid lines are obtained by using analytical solution (\ref{qsljiexi}). The corresponding asterisks are obtained through the master equation (\ref{qslshuzhi}). The atomic spontaneous emission rate is set as $\gamma=10^{-3}g$, the decay rate is assumed as $\kappa=10^{-3}g$, and the actual driving time is chosen as $\tau=1$ (in units of $g$). For the zero dissipative condition (\emph{i.e.}, $\kappa=\gamma=0$), one can find that quantum speed limit shows the almost the same behavior as with the dissipative condition. However, dissipation is bound to occur when considering the actual situation. The lines and asterisks are the quantum speed limit time based on the operator norm (red), the trace norm (green), and the Hilbert-Schmidt norm (blue), respectively. The solution of the non-Hermitian Schr\"{o}dinger agrees well with that of the master equation. The ML-type quantum speed limit bound based on the operator norm is the sharpest by comparing the three curves.

The quantum speed limit time as a function of the atom detuning $\Delta_a$ is shown in Fig. \ref{fig:2}(a), where the system parameters are set as $g=1$, $r_p=0.1$, and $\Delta_c=3g_s/\sqrt{1-\beta^2}$. As the initial atom state is the excited state $\vert e\rangle$, the excited state population of the evolved state $\langle\sigma_z\rangle$ is minimum and shows symmetry at the atom detuning near $\Delta_a=\Delta_s$, which can be concluded from the analytical expression (\ref{qsljiexi}). In the vicinity regime of $\Delta_a=\Delta_s$, the evolution of the excited state population $\langle\sigma_z\rangle$ is always negative, which will lead to a similar quantum speed limit time property. So, the ML-type quantum speed limit time reaches the maximum evolution time, \emph{i.e.}, the actual driving time. Along with the atom detuning $\Delta_a$ increased, the system will be reverted to an excited state when the parameter $\Delta_a\simeq3\Delta_s$ or $-\Delta_s$, which is similar to the Jaynes-Cummings model. However, according to Eq. (\ref{qsl}), the quantum speed limit time is determined by the initial and final state and the parameters of the system, there is no relation between the two different final states under different atom detuning $\Delta_a$ in terms of quantum speed limit. Similar mechanics can be applied to understanding quantum speed limit time variation with the cavity detuning $\Delta_c$ in Fig. \ref{fig:2}(b). The system parameters are chosen as $g=1$, $r_p=0.1$, and $\Delta_a=2g$. For the cavity detuning $\Delta_c=\Delta_a/\sqrt{1-\beta^2}$, the ML-type quantum speed limit time is the actual driving time, while the minimal evolution time is close to zero when the cavity detuning is near  $\Delta_c=4\Delta_a/\sqrt{1-\beta^2}$, or $-2\Delta_a/\sqrt{1-\beta^2}$.

The behaviors of quantum speed limit time dependent on the squeezing parameter $r_p$ is given in Fig. \ref{fig:2}(c), and the other parameters are $g=1$, $\Delta_c=3g_s/\sqrt{1-\beta^2}$, and $\Delta_a=2g$. The quantum speed limit time reaches the actual evolution time at the weakly driving laser amplitude $\Omega_p$, $\emph{i.e.}$, and the value of squeezing parameter $r_p$ is small. When the squeezing parameter $r_p$ increases to about $1.42$, the quantum speed limit time will be decreased to zero. The quantum evolution will show different characteristics with the increase of squeezing parameters. According to the effective Hamiltonian (\ref{eq:3}) in the squeezed picture, the effective coupling strength between the atom and the squeezed cavity mode is $g_s=g\cosh(r_p)$. So, similar phenomena can be observed for the evolution of the quantum speed limit time with the coupling strength $g$ in Fig. \ref{fig:2}(d) where the system parameters are $r_p=0.1$, $\Delta_c=3g_s/\sqrt{1-\beta^2}$, and $\Delta_a=2g$. It should be noted that $g$ starts from a non-zero minimum coupling strength. Otherwise, the evolution of the atom can not be affected by the cavity mode and the driving laser.

According to Fig. \ref{fig:2}, one can arrive the trend that it is more conducive to accelerate the evolution of the quantum state for the larger detuning, stronger drive and coupling strength. It is shown that the minimum evolution time can be the actual driving time or near zero when controlling the system parameters. Whether the trend for the initial excited state is suitable for other states? In the next section, we will discuss the behavior of the atom evolution when the initial state is the superposition state.

\section{The quantum speed limit time for initial superposition pure states}\label{sec4}
\begin{figure}
\centering
\includegraphics[width=1\columnwidth]{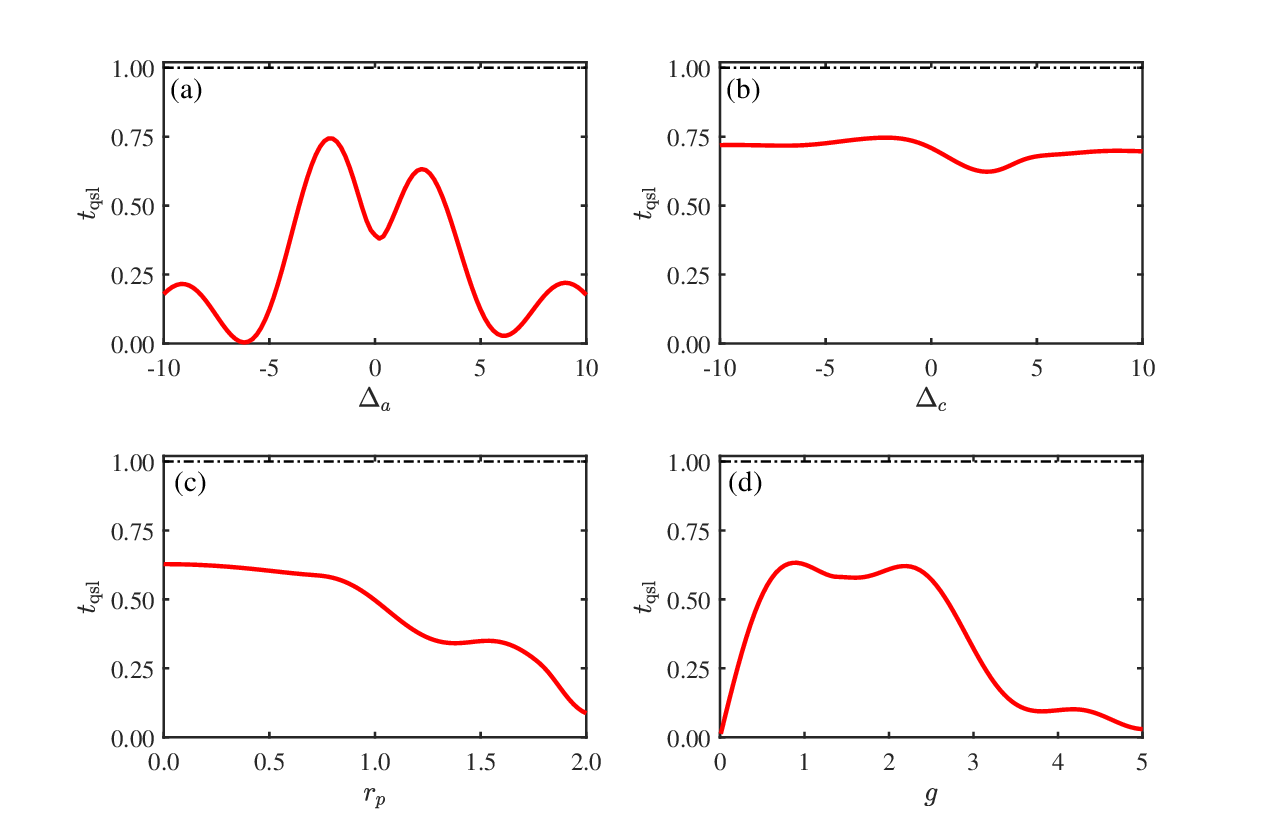}
\caption{The quantum speed limit time $t_{\text{qsl}}$ for maximal coherent state $\vert\psi(0)\rangle=(\vert e,0\rangle+\vert g,0\rangle)/\sqrt{2}$ as a function of the atom detuning $\Delta_{a}$ in Panel (a), the cavity detuning $\Delta_{c}$ in Panel (b), the squeezing parameter $r_{p}$ in Panel (c), and the coupling strength $g$ in Panel (d). The system parameters are chosen the same ones as Fig. \ref{fig:2}}\label{fig:3}
\end{figure}
\begin{figure*}
\centering
\includegraphics[width=1.6\columnwidth]{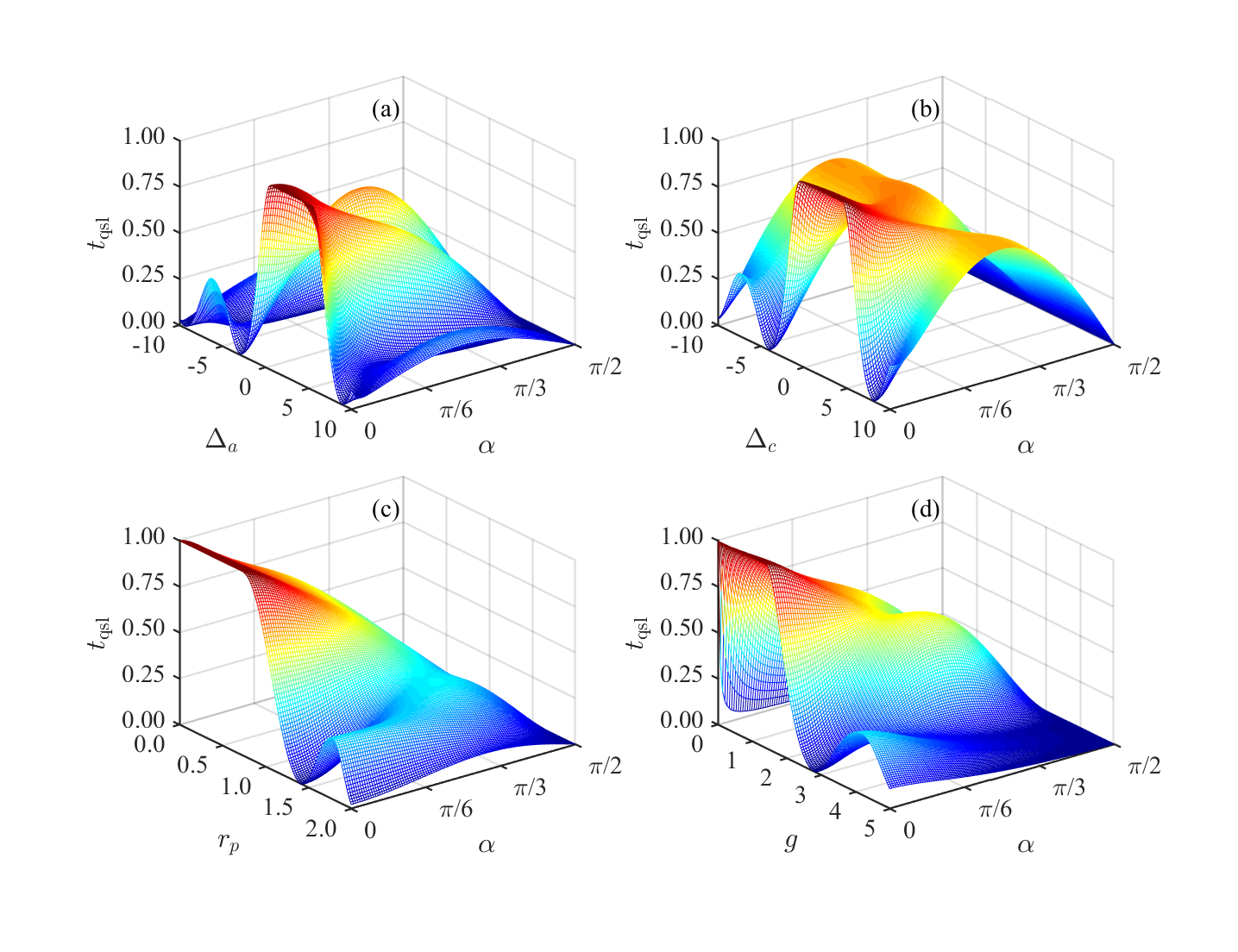}
\caption{The quantum speed limit time $t_{\text{qsl}}$ for arbitrary initial superposition state $\vert\psi(0)\rangle=\cos\alpha\vert e,0\rangle+\sin\alpha\vert g,0\rangle$ as functions of the parameter $\alpha$ and the atom detuning $\Delta_{a}$ in Panel (a), the cavity detuning $\Delta_{c}$ in Panel (b), the squeezing parameter $r_{p}$ in Panel (c), and the coupling strength $g$ in Panel (d). The system parameters are chosen the same ones as Fig. \ref{fig:2}.} \label{fig:4}
\end{figure*}
We first consider the maximal coherent initial state, $\emph{i.e.}$, $\vert\psi(0)\rangle =(\vert e,0\rangle +\vert g,0\rangle)/\sqrt{2}$. Selecting the same parameters as Fig. \ref{fig:2}, we plot the variation of ML-type quantum speed limit time with the atom detuning $\Delta_a$ in Fig. \ref{fig:3}(a), the cavity detuning $\Delta_{c}$ in Fig. \ref{fig:3}(b), the squeezing parameter $r_p$ in Fig. \ref{fig:3}(c), and the coupling strength $g$ in Fig. \ref{fig:3}(d), respectively.

In Fig. \ref{fig:3}(a), one can find that the quantum speed limit time is always shorter than the actual driving time, which means that the evolution of the state can be accelerated. Compared with Fig. \ref{fig:2}(a), a conclusion can be drawn that the quantum speed limit time is not only related to the system parameters but also dependent on the initial state \cite{Wusx2015}. The quantum speed limit time exhibits quasi-periodicity about $\Delta_a=0$ and can reach the minimum evolution time in the vicinity of $\Delta_a=-6.3g$. In Fig. \ref{fig:3}(b), we plot the quantum speed limit time variation along with the cavity detuning $\Delta_c$. In the effective Hamiltonian (\ref{eq:3}), the cavity detuning $\Delta_c$ will be shift and replaced by the squeezed cavity detuning $\Delta_s=\Delta_c\sqrt{1-\beta^2}$. Although the evolution of the quantum state can be accelerated, they have a completely different evolutionary behavior in contrast to the one in Fig. \ref{fig:3}(a).

In Fig. \ref{fig:3}(c) and (d), we plot the evolution of quantum speed limit time along with the squeezing parameter $r_p$ and the coupling strength $g$, respectively. The squeezing parameter and the quantum speed limit time have the opposite trend, \emph{i.e.}, the evolution time decreases with the increasing of $r_p$. However, the behavior of the quantum speed limit with the coupling strength $g$ is different from Fig. \ref{fig:2}(d). A reasonable explanation is that the squeezed field can the shift cavity detuning and the coupling strength between the atom and the cavity mode, and the comprehensive effects with the form of initial state govern the character of quantum speed limit time.

In the following, we will continue to investigate the quantum speed limit time for initial arbitrary superposition state $\vert\psi(0)\rangle=\cos\alpha\vert e,0\rangle+\sin\alpha\vert g,0\rangle$.  Choosing the same parameters with Fig. \ref{fig:2}, the quantum speed limit time $t_{\text{qsl}}$ are shown in Fig. \ref{fig:4} as functions of the coefficient $\alpha$ and the atom detuning $\Delta_{a}$ in Panel (a), the cavity detuning $\Delta_{c}$ in Panel (b), the squeezing parameter $r_{p}$ in Panel (c), and the coupling strength $g$ in Panel (d). The behavior of quantum speed limit time can exhibit quasi-periodicity, determined by the system parameters and the initial state. In Fig. \ref{fig:4}, one can notice that the behavior of quantum speed limit time is reduced to the conclusion in Fig. \ref{fig:2} for $\alpha=0$, $\emph{i.e.}$, the initial excited state $\vert e,0\rangle$, and the conclusion in Fig. \ref{fig:3} for $\alpha=\pi/4$, $\emph{i.e.}$, the initial maximal coherent state $(\vert e,0\rangle+\vert g,0\rangle)/\sqrt{2}$. When the initial state is the ground state $\vert g,0\rangle$, $\emph{i.e.}$, $\alpha=\pi/2$, the state does not evolve along with the time, so zero quantum speed limit time is reasonable. In Fig. \ref{fig:4}(c) and (d), the trend of quantum speed limit time is descending with the increasing of $r_p$ or $g$, which indicates that the higher effective coupling strength, the faster the quantum state evolution, which fits the intuition in the field of quantum optics.

\section{Discussion and conclusion}\label{sec5}
Now we discuss the feasibility of this model in the experiment. High fineness Fabry-Perot resonator is a single-mode optical cavity, and Cesium or Rubidium can be selected as experimental atoms. The squeezed cavity mode is generated by second-order nonlinear medium pumping \cite{Boca2004PRL}, in which the squeezing parameter can be adjusted by the amplitude and relative phase of the laser. Additional broadband squeezed vacuum field with bandwidth from MHz to GHz is generated via periodically poled potassium titanyl phosphate (PPKTP) crystal \cite{Serikawa2016}. Choosing the following parameters: $g=3\times2\pi \text{ MHz}$, $\kappa=3\times2\pi \text{ kHz}$, $\gamma=3\times2\pi \text{ kHz}$, $r_{p}=r_{e}=0.1$, $\Delta_{a}=2g$, $\Delta_{c}=3g\cosh(r_{p})/\sqrt{1-\beta^2}$, $\theta_{e}=\pi$, $\theta_{p}=0$, our theoretical model can be verified based on the current experimental techniques.

In summary, we have studied the quantum speed limit time for an atom trapped in a fineness Fabry-Perot resonator with a second-order medium driven by a classical coherent laser. The effective Hamiltonian is arrived at using the Bogoliubov squeezing transformation, and analytical expression of the reduced density matrix is obtained for the initial excited state $\vert e,0\rangle$. The quantum speed limit time based on the analytical solution agrees with the Lindblad master equation. The quantum speed limit can reach the actual driving time when the evolution is not accelerated. At the same time, it can approach zero when the state evolves back to the initial state, which is determined by the system parameters. From the perspective of quantum speed limit, it is more conducive to accelerate the evolution of the quantum state for the large detuning, strong drive and coupling strength. We also investigate the ML-type quantum speed limit time for the initial maximal coherent state  $(\vert e,0\rangle +\vert g,0\rangle)/\sqrt{2}$ and the initial arbitrary superposition state  $\cos\alpha\vert e,0\rangle+\sin\alpha\vert g,0\rangle$. The quantum speed limit time exhibits quasi-periodicity and abundant phenomena, and it is not only dependent on the system parameters but also determined by the initial state, which the form of initial state has more influence on the quantum speed limit time. We can utilize the initial state and the specific system parameters to control the evolution of the quantum system to the field of quantum optics, cavity quantum electrodynamics, and pave a novel view to explanation the experimental phenomenon for a single atom interacting with the impact of ambient noise in the squeezed cavity environment.

\section*{ACKNOWLEDGMENT}
This work was supported by the National Natural Science Foundation of China (Grant No. 12175029), Fundamental Research Program of Shanxi Province (Grant No. 20210302123063).

\section*{Appendix: Derivation of the master equation (\ref{qslshuzhi})}
The master equation of the whole system is
\begin{align}
\dot{\rho}(t)= & i[\rho(t),H_R]-\frac{1}{2}\{\mathcal{L}(L_{\sigma_-})\rho(t)+\mathcal{L}(L_a)\rho(t)\},\label{eq:11}
\end{align}
where $H_R$ is the Hamiltonian (\ref{eq:2}), $L_{\sigma_-}=\sqrt{\gamma}\sigma_{-}$ describes the atomic spontaneous emission with rate $\gamma$, and $L_a=\sqrt{\kappa}a$ describes the bare cavity decay with rate $\kappa$. Using the Bogoliubov squeezing transformation $a=a_{s}\cosh(r_{p})-e^{i\theta_{p}}\sinh(r_{p})a_{s}^{\dagger}$,
the master equation (\ref{eq:11}) can be rewritten as
\begin{align}
\dot{\rho}(t)= & i[\rho(t),H]-\frac{1}{2}\{\mathcal{L}(L_{\sigma_-})\rho(t)\notag \\
&+(N+1)\mathcal{L}(L_{as})\rho(t)+N\mathcal{L}(L_{as}^{\dagger})\rho(t)\notag \\
&-M\mathcal{L}'(L_{as}^{\dagger})\rho(t)-M^{*}\mathcal{L}'(L_{as})\rho(t)\},
\end{align}
where $H$ is the effective Hamiltonian (\ref{eq:3}), and $L_{as}=\sqrt{\kappa}a_{s}$ is the Lindblad operator describing the decay of the squeezed-cavity mode. $N=\sinh^{2}(r_{p})$ and $M=e^{-i\theta_p}\cosh(r_{p})\sinh(r_{p})$ describe the effective thermal noise and two-photon correlation \cite{scully1997,Breuer book}, respectively. The superoperator $\mathcal{L}(o)$ and $\mathcal{L}'(o\text{)}$ are defined as
\begin{align}
\mathcal{L}(o)\rho(t)=&o^{\dagger}o\rho(t)-2o\rho(t)o^{\dagger}+\rho(t)o^{\dagger}o,\notag \\
\mathcal{L}'(o)\rho(t)=&oo\rho(t)-2o\rho(t)o+\rho(t)oo.
\end{align}

When the driving laser is applied to the nonlinear medium $\chi^{(2)}$, the squeezed noise will be introduced synchronously. However, external broadband squeezed vacuum field can suppress the additional noise. When the cavity mode is coupled with the squeezed vacuum reservoir with the squeezing parameter $r_{e}$ and the reference phase $\theta_{e}$, the master equation of the trapped atom and the cavity mode is given as follows:
\begin{align}
\dot{\rho}(t)= & i[\rho(t),H_R]-\frac{1}{2}\{\mathcal{L}(L_{\sigma_-})\rho(t)\notag \\
 & +(N'+1)\mathcal{L}(L_{a})\rho(t)+N'\mathcal{L}(L_{a}^{\dagger})\rho(t)\notag \\
 & -M'\mathcal{L}'(L_{a}^{\dagger})\rho(t)-M'^{*}\mathcal{L}'(L_{a})\rho(t)\},\label{eq:14}
\end{align}
where $N'=\sinh^{2}(r_{e}),~M'=e^{i\theta_{e}}\cosh(r_{e})\sinh(r_{e})$ are parameters that describe the squeezed vacuum reservoir. Following is the process that obtain the master equation (\ref{eq:11}), taking the Bogoliubov squeezing transformation $a=a_{s}\cosh(r_{p})-e^{i\theta_{p}}\sinh(r_{p})a_{s}^{\dagger}$ into the Eq. (\ref{eq:14}), the master equation in the squeezed cavity mode picture is obtained as follows
\begin{align}
\dot{\rho}(t)= & i[\rho(t),H]-\frac{1}{2}\{\mathcal{L}(L_{\sigma_-})\rho(t)\notag \\
 & +(N_{s}+1)\mathcal{L}(L_{as})\rho(t)+N_{s}\mathcal{L}(L_{as}^{\dagger})\rho(t)\notag \\
 & -M_{s}\mathcal{L}'(L_{as}^{\dagger})\rho(t)-M_{s}^{*}\mathcal{L}'(L_{as})\rho(t)\},
\end{align}
where $N_{s}$ and $M_{s}$ are given by
\begin{align}
N_{s}= & \sinh^{2}(r_{e})\cosh(2r_{p})+\sinh^{2}(r_{p})\notag \\
 & +\frac{1}{2}\sinh(2r_{p})\sinh(2r_{e})\cos(\theta_{e}+\theta_{p}),\notag \\
M_{s}= & -e^{-i\theta_{p}}\{\frac{1}{2}\sinh(2r_{p})\cosh(2r_{e})+\frac{1}{2}\sinh(2r_{e})\notag\\
 & [e^{i(\theta_{e}+\theta_{p})}\cosh^{2}(r_{p})+e^{-i(\theta_{e}+\theta_{p})}\sinh^{2}(r_{p})]\}.
\end{align}
When the squeezed parameters are chosen appropriately, \emph{i.e.}, $r_{e}=r_{p}$, $\theta_{e}+\theta_{p}=\pi$, the parameters $N_{s}$ and $M_{s}$ can be reduced to zero. The influence of noise caused by the driving laser $\Omega_p$ can be completely suppressed. The dynamical evolution of the quantum system can be described by the master equation (\ref{qslshuzhi}).


\begin{thebibliography}{10}
\bibitem{Mandelstam1945} Mandelstam L and Tamm I 1945 \emph{J. Phys.} (USSR) \textbf{9} 249
\bibitem{Margolus1998} Margolus N and Levitin L B 1998 \emph{Phys. D} \textbf{120} 188
\bibitem{Anandan1990PRL} Anandan J and Aharonov Y 1990 \emph{Phys. Rev. Lett.} \textbf{65} 1697
\bibitem{Fleming1973} Fleming G N 1973 \emph{Nuovo Cimento} \textbf{16} 232
\bibitem{Bhattacharyya1983} Bhattacharyya K 1983 \emph{J. Phys. A: Math. Gen.} \textbf{16} 2993
\bibitem{Vaidman1992} Vaidman L 1992 \emph{Am. J. Phys.} \textbf{60} 182
\bibitem{Caneva2009PRL} Caneva T, Murphy M, Calarco T, Fazio R, Montangero S, Giovannetti V and Santoro G E 2009 \emph{Phys. Rev. Lett.} \textbf{103} 240501
\bibitem{Hegerfeldt2013PRL} Hegerfeldt G C 2013 \emph{Phys. Rev. Lett.} \textbf{111} 260501
\bibitem{Campbell2017PRL} Campbell S and Deffner S 2017 \emph{Phys. Rev. Lett.} \textbf{118} 100601
\bibitem{Frank2016} van Frank S, Bonneau M, Schmiedmayer J,  Hild S, Gross C, Cheneau M, Bloch I, Pichler T, Negretti A, Calarco T and Montangero S 2016 \emph{Sci. Rep.} \textbf{6} 34187
\bibitem{Giovannetti2011} Giovannetti V, Lloyd S and Maccone L 2011 \emph{Nat. Phot.} \textbf{5} 222
\bibitem{Chin2012PRL} Chin A W, Huelga S F and Plenio M B 2012 \emph{Phys. Rev. Lett.} \textbf{109} 233601
\bibitem{Deffner2010PRL} Deffner S and Lutz E 2010 \emph{Phys. Rev. Lett.} \textbf{105} 170402
\bibitem{Greiner2002Nature} Greiner M, Mandel O, Esslinger T, H\"{a}nsch T W and Bloch I 2002 \emph{Nature} (London) \textbf{415} 39
\bibitem{Sachdev book} Sachdev S 2011 \emph{Quantum Phase Transitions} (Cambridge: Cambridge University Press )
\bibitem{Deffner2017JPA} Deffner S and Campbell S 2017 \emph{J. Phys. A: Math. Theor.} \textbf{50} 453001
\bibitem{del Campo2013PRL} del Campo A, Egusquiza I L, Plenio M B and Huelga S F 2013 \emph{Phys. Rev. Lett.} \textbf{110} 050403
\bibitem{Taddei2013PRL} Taddei M M, Escher B M, Davidovich L and de Matos Filho R L 2013 \emph{Phys. Rev. Lett.} \textbf{110} 050402
\bibitem{Deffner2013PRL} Deffner S and Lutz E 2013 \emph{Phys. Rev. Lett.} \textbf{111} 010402
\bibitem{Zhang2014} Zhang Y J, Han W, Xia Y J, Cao J P and Fan H 2014 \emph{Sci. Rep.} \textbf{4} 4890
\bibitem{Xu2014} Xu Z Y, Luo S L, Yang W L, Liu C and Zhu S Q 2014 \emph{Phys. Rev. A} \textbf{89} 012307
\bibitem{wusx2018} Wu S X and Yu C S 2018 \emph{Phys. Rev. A} \textbf{98} 042132
\bibitem{wusx2020} Wu S X and Yu C S 2020 \emph{Sci. Rep.} \textbf{10} 5500
\bibitem{Zhang2018} Zhang L, Sun Y and Luo S L 2018 \emph{Phys. Lett. A} \textbf{382} 2599-2604
\bibitem{sun2015} Sun Z, Liu J, Ma J and Wang X 2015 \emph{Sci. Rep.} \textbf{5} 8444
\bibitem{Liu2016} Liu H B, Yang W L, An J H and Xu Z Y 2016 \emph{Phys. Rev. A} \textbf{93} 020105
\bibitem{song2016} Song Y J, Kuang L M and Tan Q S 2016 \emph{Quantum Inf Process} \textbf{15} 2325
\bibitem{wei2016} Wei Y B, Zou J, Wang Z M, Shao B and Li H 2016 \emph{Phys. Lett. A} \textbf{380} 397
\bibitem{cai2017} Cai X J and Zheng Y J 2017 \emph{Phys. Rev. A} \textbf{95} 052104
\bibitem{huang2022cpb} Huang J H, Qin L G, Chen G L, Hu L Y and Liu F Y 2022 \emph{Chin. Phys. B} \textbf{31} 110307
\bibitem{lin2022cpb} Lin Z Y, liu T, Li Z L, Zhang Y H and Lan K 2022 \emph{Chin. Phys. B} \textbf{31} 070307
\bibitem{zhang2015} Zhang Y J, Han W, Xia Y J, Cao J P and Fan H 2015 \emph{Phys. Rev. A} \textbf{91} 032112
\bibitem{xu2019} Xu K, Zhang G F and Liu W M 2019 \emph{Phys. Rev. A} \textbf{100} 052305
\bibitem{Yu2018} Yu M, Fang M F and Zou H M 2018 \emph{Chin. Phys. B} \textbf{27} 010303
\bibitem{O'Connor2021} O'Connor E, Guarnieri G and Campbell S 2021 \emph{Phys. Rev. A} \textbf{103} 022210
\bibitem{Mondal2016} Mondal D, Datta C and Sazim S 2016 \emph{Phys. Lett. A} \textbf{380} 689
\bibitem{Jones2010} Jones P J and Kok P 2010 \emph{Phys. Rev. A} \textbf{82} 022107
\bibitem{Uzdin} Uzdin R and Kosloff R 2016 \emph{Europhys. Lett.} \textbf{115} 40003
\bibitem{Teittinen2019} Teittinen J, Lyyra H and Maniscalco S 2019 \emph{New J. Phys.} \textbf{21} 123041
\bibitem{Ektesabi2017} Ektesabi A, Behzadi N and Faizi E 2017 \emph{Phys. Rev. A} \textbf{95} 022115
\bibitem{Deffner2017} Deffner S 2017 \emph{New J. Phys.} \textbf{19} 103018
\bibitem{Diaz2020} D\`{\i}az V A A, Martikyan V, Glaser S J and Sugny D 2020 \emph{Phys. Rev. A} \textbf{102} 033104
\bibitem{Burgarth2022} Burgarth D, Borggaard J and Zimbor\'{a}s Z 2022 \emph{Phys. Rev. A} \textbf{105} 042402
\bibitem{cheng2022} Cheng W W, Li B, Gong L Y and Zhao S M 2022 \emph{Phys. A} \textbf{597} 127242
\bibitem{Mohan2022} Mohan B, Das S and Pati A K 2022 \emph{New J. Phys.} \textbf{24} 065003
\bibitem{Ness2022PRL} Ness G, Alberti A and Sagi Y 2022 \emph{Phys. Rev. Lett.} \textbf{129} 140403
\bibitem{Liu2017} Liu X, Wu W and Wang C 2017 \emph{Phys. Rev. A} \textbf{95} 052118
\bibitem{Funo2019} Funo K, Shiraishi N and Saito K 2019 \emph{New J. Phys.} \textbf{21} 013006
\bibitem{Hu2020} Hu X H, Sun S N and Zheng Y J 2020 \emph{Phys. Rev. A} \textbf{101} 042107
\bibitem{Vu2021PRL} Vu T V and Hasegawa Y 2021 \emph{Phys. Rev. Lett.} \textbf{127} 190601
\bibitem{liu2015} Liu C, Xu Z Y and Zhu S Q 2015 \emph{Phys. Rev. A} \textbf{91} 022102
\bibitem{Levitin2009PRL} Levitin L B and Toffoli T 2009 \emph{Phys. Rev. Lett.} \textbf{103} 160502
\bibitem{Campaioli2018PRL} Campaioli F, Pollock F A, Binder F C and Modi K 2018 \emph{Phys. Rev. Lett.} \textbf{120} 060409
\bibitem{Wusx2015} Wu S X, Zhang Y, Yu C S and Song H S 2015 \emph{J. Phys. A: Math. Theor.} \textbf{48} 045301
\bibitem{Mirkin2016} Mirkin N, Toscano F and Wisniacki D A 2016 \emph{Phys. Rev. A} \textbf{94} 052125
\bibitem{Marvian2016} Marvian I, Spekkens R W and Zanardi P 2016 \emph{Phys. Rev. A} \textbf{93} 052331
\bibitem{Poggi2019} Poggi P M 2019 \emph{Phys. Rev. A} \textbf{99} 042116
\bibitem{Du2021} Du K Y, Ma Y J, Wu S X and Yu C S 2021 \emph{Chin. Phys. B} \textbf{30} 090308
\bibitem{Tian2019} Tian C, Lu X, Zhang Y J and Xia Y J 2019 \emph{Acta Phys. Sin.} \textbf{68} 150301
\bibitem{Shiraishi2018PRL} Shiraishi N, Funo K and Saito K 2018 \emph{Phys. Rev. Lett.} \textbf{121} 070601
\bibitem{Okuyama2018PRL} Okuyama M and Ohzeki M 2018 \emph{Phys. Rev. Lett.} \textbf{120} 070402
\bibitem{Shanahan2018PRL} Shanahan B, Chenu A, Margolus N and del Campo A 2018 \emph{Phys. Rev. Lett.} \textbf{120} 070401
\bibitem{wucpb2020} Wu S X and Yu C S 2020 \emph{Chin. Phys. B} \textbf{29} 050302
\bibitem{Garcia-Pintos2022} Garc\'{\i}a-Pintos L P, Nicholson S B, Green J R, del Campo A and Gorshkov A V 2022 \emph{Phys. Rev. X} \textbf{12} 011038
\bibitem{Liu2021} Liu J, Miao Z B, Fu L B and Wang X G 2021 \emph{Phys. Rev. A} \textbf{104} 052432
\bibitem{Shao2020} Shao Y Y, Liu B, Zhang M, Yuan H D and Liu J 2020 \emph{Phys. Rev. Res.} \textbf{2} 023299
\bibitem{Sun2019PRL} Sun S N and Zheng Y J 2019 \emph{Phys. Rev. Lett.} \textbf{123} 180403
\bibitem{Sun2021PRL} Sun S N, Peng Y G, Hu X H and Zheng Y J 2021 \emph{Phys. Rev. Lett.} \textbf{127} 100404
\bibitem{scully1997} Scully M O and Zubairy M S 1997 \emph{Quantum Optics} (Cambridge: Cambridge University Press)
\bibitem{purcell1946} Purcell E M 1946 \emph{Phys. Rev.} \textbf{69} 681
\bibitem{Bloembergen1954} Bloembergen N and Pound R V 1954 \emph{Phys. Rev.} \textbf{95} 8
\bibitem{Birnbaum2005} Birnbaum K M, Boca A, Miller R, Boozer A D, Northup T E and Kimble H J 2005 \emph{Nature} \textbf{436} 87
\bibitem{Hamsen2017} Hamsen C, Tolazzi K N, Wilk T and Rempe G 2017 \emph{Phys.Rev.Lett.} \textbf{118} 133604
\bibitem{Yang2019PRL} Yang P F, Xia X W, He H, Li S K, Han X, Zhang P, Li G, Zhang P F, Xu J P, Yang Y P and Zhang T C 2019 \emph{Phys. Rev. Lett.} \textbf{123} 233604
\bibitem{Yang2019} Yang P F, Li M, Han X, He H, Li G, Zou C L, Zhang P F and Zhang T C 2019 arXiv: 1911.10300
\bibitem{Lv2015PRL} L\"{u} X Y, Wu Y, Johansson J R, Jing H, Zhang J and Nori F 2015 \emph{Phys. Rev. Lett.} \textbf{114} 093602
\bibitem{Qin2018PRL} Qin W, Miranowicz A, Li P B, L\"{u} X Y, You J Q and Nori F 2018 \emph{Phys. Rev. Lett.} \textbf{120} 093601
\bibitem{Wang2019} Wang Y, Li C, Sampuli E M, Song J, Jiang Y Y and Xia Y 2019 \emph{Phy. Rev. A} \textbf{99} 023833
\bibitem{Boca2004PRL} Boca A, Miller R, Birnbaum K M, Boozer A D, McKeever J and Kimble H J 2004  \emph{Phys. Rev. Lett.} \textbf{93} 233603
\bibitem{Serikawa2016} Serikawa T, ichi Yoshikawa J, Makino K and Frusawa A 2016 \emph{Opt. Express} \textbf{24} 28383
\bibitem{Breuer book} Breuer H P and Petruccione F 2002 \emph{The Theory of Open Quantum Systems} (New York: Oxford University Press)
\end{thebibliography}
\end{document}